\documentclass{aa}

\usepackage{graphicx}
\usepackage{txfonts}
\usepackage{hyperref}
\usepackage{orcidlink}

\begin{document} 

%\nolinenumbers
\title{Propeller effect in action: Unveiling quenched accretion in the transient X-ray pulsar \mbox{4U 0115+63}}

\author{%\small 
Hua~Xiao \inst{\ref{in:UTU}}\orcidlink{0009-0004-1288-4912}
\and Sergey~S.~Tsygankov \inst{\ref{in:UTU}}\orcidlink{0000-0002-9679-0793}
\and Valery~F.~Suleimanov \inst{\ref{in:Tub}}\orcidlink{0000-0003-3733-7267}
\and Alexander~A.~Mushtukov \inst{\ref{in:Oxford}}\orcidlink{0000-0003-2306-419X}
\and Long~Ji \inst{\ref{in:sysu},\ref{in:sysu_csst}}\orcidlink{0000-0001-9599-7285}
\and Juri~Poutanen \inst{\ref{in:UTU}}\orcidlink{0000-0002-0983-0049}
          }
          
\institute{Department of Physics and Astronomy, FI-20014 University of Turku,  Finland \label{in:UTU} \\ \email{hua.h.xiao@utu.fi}
\and 
Institut f\"ur Astronomie und Astrophysik, Universit\"at T\"ubingen, Sand 1, D-72076 T\"ubingen, Germany \label{in:Tub}
\and 
Astrophysics, Department of Physics, University of Oxford, Denys Wilkinson Building, Keble Road, Oxford OX1 3RH, UK \label{in:Oxford}
\and
School of Physics and Astronomy, Sun Yat-sen University, Zhuhai, 519082, People’s Republic of China \label{in:sysu}
\and
CSST Science Center for the Guangdong-Hong Kong-Macau Greater Bay Area, DaXue Road 2, 519082, Zhuhai, People's Republic of China \label{in:sysu_csst}
}
          
\titlerunning{Propeller effect in action in the transient X-ray pulsar \mbox{4U 0115+63}}
\authorrunning{Xiao et al.}

\abstract 
{The Be/X-ray pulsar \mbox{4U 0115+63} underwent a type II outburst in 2023. After the outburst, similar to the outbursts in 2015 and 2017, the source decayed into a quiescent state. Two out of three {\it XMM-Newton} observations conducted after the 2023 outburst confirmed the source to be in a low-luminosity state at a level of $L_{\rm X} \sim 10^{33}\,\rm erg\,s^{-1}$. X-ray pulsations were detected at $\approx$0.277\,Hz in both observations with a pulsed fraction exceeding 50\%. The power density spectra show no significant low-frequency red noise in either observation, suggesting that the radiation is not driven by accretion. The energy spectra in this state can be described by a single blackbody component, with an emitting area smaller than the typical size of the polar caps during the accretion phase. Based on the timing and spectral properties, we suggest that the propeller effect is active during the quiescent state, resulting in a total quenching of accretion. We discuss possible mechanisms for the generation of pulsations in this regime and consider the scenario of neutron star crust cooling.}


\maketitle

\section{Introduction} 
\label{sec:intro}

Transient X-ray pulsars (XRPs) in Be X-ray binaries (BeXRPs) host a highly magnetized neutron star (NS) with a magnetic field on the order of $B \sim 10^{12}$--$10^{13}$\,G. BeXRPs emit X-rays when accreting matter from a decretion disk that episodically forms around a rapidly rotating Be-type companion star. As the accreted material approaches the NS magnetosphere, it couples to the magnetic field lines and is funneled onto the magnetic poles. This process releases gravitational potential energy, which is radiated as X-rays. The anisotropic nature of the accretion flow leads to pulsed X-ray emission modulated by the NS rotation \citep[for a review, see][]{MushtukovTsygankov2024}.
These systems usually have significantly eccentric orbits. Near periastron, the increase in the mass accretion rate can trigger a so-called type I outburst, with X-ray luminosities reaching up to $L_{\rm X} \sim 10^{37}\,\mathrm{erg\,s^{-1}}$. In other cases, BeXRPs can reach luminosities above the Eddington limit for a NS during giant (type II) outbursts. The physical mechanism behind type II outbursts remains uncertain \citep[see the review by][]{Reig2011}.

The temporal evolution of the observed X-ray flux from transient XRPs is governed to a large extent by the interaction of matter with the magnetic field. For instance, for the systems with fast-spinning NSs (typical spin period $P_{\rm spin} \lesssim 10\,\rm s$), at the end of the outburst, when the mass accretion rate decreases to a threshold value, the accretion is expected to be inhibited by the centrifugal barrier -- the so-called propeller regime \citep{Illarionov1975}. The threshold mass accretion rate corresponding to the transition to the propeller regime corresponds to the equality of the magnetospheric and corotation radii of the NS and can be expressed as \citep[e.g.,][]{Stella1986}
$$L_{\rm prop} \simeq 4 \times 10^{37}\,k^{7/2}\, B^{2}_{12}\,P_{\rm spin}^{-7/3}\,M^{-2/3}_{1.4}\,R^{5}_6\,\rm erg\,s^{-1},$$
where $B_{12}$ is the magnetic field strength in units of $10^{12} \, \text{G}$, $P_{\rm spin}$ is the spin period in seconds, $M_{1.4}$ is the NS mass in units of $1.4 \, M_\odot$, $R_6$ is the radius in $10^6 \, \text{cm}$, and $k$ is a geometric factor of order unity. Below this luminosity threshold, accretion onto the NS is expected to cease abruptly, resulting in a sharp decline in observed flux and a transition to a low-luminosity quiescent state. In several XRPs, dramatic luminosity evolutions have been observed in the final stages of their outbursts and are considered to result from the propeller effect, for example SXP~4.78 \citep{Semena2019}, GRO~J1750$-$27 \citep{Lutovinov2019}, \mbox{4U 0115+63}, and V~0332+53 \citep{Stella1986, Campana2001, Tsygankov2016}.

At the same time, most transient XRPs observed in quiescence exhibit a nonzero flux in the X-ray band, with a typical luminosity of \( L_{0.5-10\,\mathrm{keV}} \lesssim 10^{34}\,\mathrm{erg\,s^{-1}} \) \citep{Tsygankov2017}. Moreover, some sources still display pulsations during this state (e.g., 4U~1145$-$619, \citealt{Mukherjee2005}; 1A~1118$-$615, \citealt{Rutledge2007}; 1A~0535+262, \citealt{Mukherjee2005}; \mbox{and 4U 0115+63}, \citealt{Rouco2017,Rouco2020}; see also Table 1 in \citealt{Tsygankov2017} for a complete source list). The reason for this phenomenon remains unknown, with two main hypotheses proposed. The first is the leakage of matter through the centrifugal barrier along magnetic field lines, which leads to the formation of hot spots \citep[see, e.g.,][]{Elsner1977, Ikhsanov2001, Lii2014, Orlandini2004, Mukherjee2005}. The second is thermal emission from a cooling, accretion-heated NS, in which more heat from the solid crustal layer is conducted to the magnetic poles due to anisotropic heat transport governed by the magnetic field lines \citep{Rouco2017}.

Therefore, the possibility of accretion onto the NS surface in the propeller regime remains viable and has not been definitively ruled out by observational data. Determining the origin of the emission in the quiescent state of a transient XRP requires high-quality data obtained during this phase. Such data were acquired with the \textit{XMM-Newton} observatory for the well-studied pulsar \mbox{4U 0115+63} following its 2023 outburst.

\mbox{4U 0115+63} is a typical transient BeXRP that was discovered by the \textit{Uhuru} satellite \citep{Giacconi1972} and identified as having a B0.2Ve spectral type companion star \citep{Johns1978}. The NS has a magnetic field strength  $B\approx 1.3 \times 10^{12}\,\mathrm{G}$ as measured from the cyclotron line energy \citep{Heindl2000} and a spin period  $P_{\mathrm{spin}}\approx 3.61\,\mathrm{s}$ \citep{Cominsky1978}. It orbits its companion every 24.3~d with an eccentricity of approximately 0.34 \citep{Rappaport1978}. The distance to the source is estimated to be $\approx 5.0\,\mathrm{kpc}$, based on the \textit{Gaia} data \citep{Neumann2023}.

Over the last decade, \mbox{4U 0115+63} has experienced several giant type II outbursts. \citet{Tsygankov2016} discovered a rapid drop in flux following the 2015 outburst and suggested that the source had transitioned into the propeller regime. The authors estimated the threshold luminosity for this transition to be $(1.4 \pm 0.4) \times 10^{36}\,\mathrm{erg\,s^{-1}}$. Based on follow-up observations using \textit{Swift} and \textit{XMM-Newton} at the end of this and the following outbursts, the source was found to be in a low-luminosity state with $L_{\rm 0.5-10~keV} \sim 10^{33}\,\mathrm{erg\,s^{-1}}$ \citep{Wijnands2016, Rouco2017, Rouco2020}. X-ray pulsations were detected in the \textit{XMM-Newton} observations during this state \citep{Rouco2017, Rouco2020}.

In this study we performed timing and spectral analysis of the  deep \textit{XMM-Newton} observations  of \mbox{4U 0115+63} obtained after the 2023 giant outburst to constrain the origin of its pulsating emission in the quiescent state.

\begin{table*}
\centering
\renewcommand\arraystretch{1.3}
\caption{Details of the {\it XMM-Newton} observations of \mbox{4U 0115+63} used in this paper.}
\begin{tabular}{ccccccc}
\hline
\hline
  Obs No. & ObsID & Time & PN exposure time & Count rate\tablefootmark{a} & Threshold rate\tablefootmark{b}  \\
  &  & (MJD)  & (ks) & ($\,\rm count\,s^{-1}$) &  (count\,s$^{-1}$) \\ \hline
  Obs1 &  0883350401 & 60138.72--60138.84  & $\sim$6.2 & 0.49$\pm$0.04 & $\geq 0.55$, $\geq 0.2$, $\geq 0.25$\\
  Obs2 &  0883350301 & 60160.55--60160.78  & $\sim$5.9 & 0.43$\pm$0.04 &  $\geq 0.45$, $\geq 0.15$, $\geq 0.2$ \\
  Obs3 &  0883351701 & 60198.44--60198.67  & $\sim$13.3 & 7.85$\pm$0.09 & $\geq 1.0$, $\geq 0.2$, $\geq 0.25$ \\
  Obs2016 & 0790180301 & 57435.30--57435.91  & $\sim$21.3 & 0.32$\pm$0.04 & $\geq 0.55$, $\geq 0.2$, $\geq 0.25$ \\
        \hline
\end{tabular}
\tablefoot{\tablefoottext{a}{The average count rates are obtained from PN light curves in 0.3--10\,keV band and the errors are 1$\sigma$.} 
\tablefoottext{b}{The threshold rate are shown in the following order: PN, MOS1 and MOS2. All observations were conducted in Full-Frame science mode.}} 
\label{table:obs}
\end{table*}

\section{Observations and data reduction} 
\label{sec:data}

\begin{figure}
\centering
\includegraphics[width=1.0\linewidth]{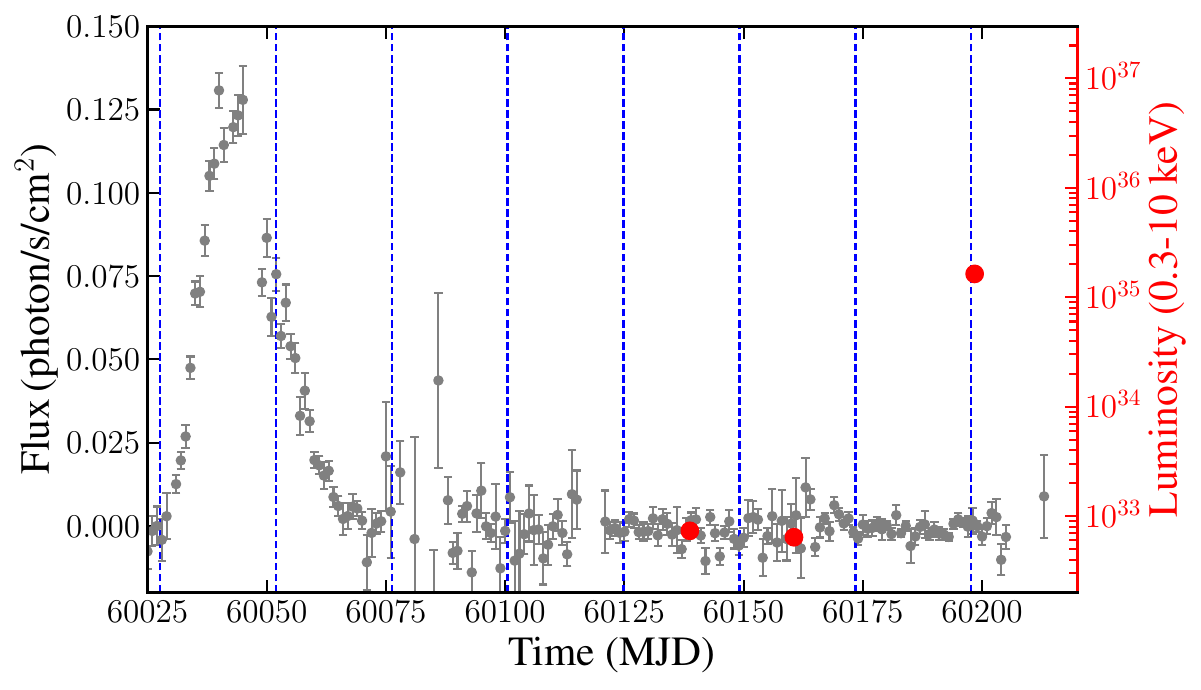}
\caption{Light curves of \mbox{4U 0115+63} during and after the 2023 giant outburst observed with {\it Swift}/BAT (gray) and {\it XMM-Newton} (red). The luminosities of {\it XMM-Newton} observations in the 0.3--10\,keV band are $(7.3^{+0.5}_{-0.4}) \times 10^{32}\,\rm erg\,s^{-1}$, $(6.4^{+0.4}_{-0.7}) \times 10^{32}\,\rm erg\,s^{-1}$, and $(1.65\pm0.02) \times 10^{35}\,\rm erg\,s^{-1}$ for Obs1, Obs2, and Obs3, respectively, obtained by spectral analysis described in Sect.~\ref{sec:spec} using a distance of 5.0\,kpc. The vertical  dashed blue lines represent the times of periastron passages determined by adopting the ephemeris of \cite{Raichur2010}.}
\label{fig:lc}
\end{figure}

In March 2023, \mbox{4U 0115+63} exhibited a new type II outburst, followed by a decay into a low-luminosity (quiescent) state. \textit{XMM-Newton} observed the source multiple times, and we used the suitable data from the \textit{XMM} European Photon Imaging Camera (EPIC), specifically the EPIC-PN (hereafter PN) 
%(p-n-junction) 
and EPIC-MOS (hereafter MOS)
%(metal oxide semi-conductor) 
detectors. Observations were conducted on 2023 July 13 (MJD 60138, hereafter Obs1), August 4 (MJD 60160, Obs2), and September 11 (MJD 60198, Obs3). We also included an observation conducted on 2016 February 17 (MJD 57435; Obs2016), when the source was in the quiescence (see Table~\ref{table:obs} and Fig.~\ref{fig:lc} for details).

Following the official SAS Science threads,\footnote{\url{https://www.cosmos.esa.int/web/xmm-newton/sas-threads}} the data reduction was performed by using the Science Analysis System (SAS v22.1.0)\footnote{\url{https://www.cosmos.esa.int/web/xmm-newton/sas}} software with the latest calibration datasets. We used the {\tt epproc} and {\tt emproc} tasks to produce calibrated and concatenated event lists from the PN and MOS cameras, respectively. To remove high background flaring activity from the observations, the single-event ({\tt PATTERN==0}) light curves in the 10--12\,keV band for the PN camera and $>$10\,keV for the MOS cameras were extracted; then we used the threshold rates for a steady low background (see Table~\ref{table:obs}) to get the good time intervals of each observation. Due to the very low count rates of the source during the {\it XMM-Newton} observations, Obs1, Obs2, and Obs2016 were not affected by pile-up. For these observations, the source region was identified as a circular region with a 20\arcsec\ radius, and the background was extracted from a nearby, source-free circular region with a 50\arcsec\ radius on the same CCD chip. For Obs3, to mitigate pile-up effects, we used an annular source region centered on the source position, with an inner radius of 10\arcsec\ and an outer radius of 20\arcsec. We extracted the single and double events ({\tt PATTERN $\leq$ 4}) for the PN camera and the events between single and quadruple ({\tt PATTERN $\leq$ 12}) for the MOS cameras. The barycentric correction was performed by using {\tt barycen} task.

For the timing analysis, we used the {\tt epiclccorr} task to obtain net light curves. We used the {\tt powspec} tool to get the power density spectra (PDSs) and folded the pulse profiles using the {\tt efold} tool from the HEASoft v6.35.1 package.\footnote{\url{https://heasarc.gsfc.nasa.gov/docs/software/lheasoft}}
The background spectra were scaled using the {\tt  backscale} task. The response matrix and the ancillary response files were generated using {\tt rmfgen} and {\tt arfgen}, respectively. We used the {\tt epicspeccombine} task to combine the spectra of three EPIC cameras in order to improve statistics. Due to the very few counts in the spectra, we re-binned the spectra with {\tt grppha} tool to make sure each bin has a minimum count of 5. The spectral analysis was performed by using the X-ray spectral fitting package ({\sc xspec}) version 12.15.0 \citep{Arn96}. We used C-statistics for the statistics of fitting. All uncertainties in this paper for spectral analysis correspond to a confidence level of 90\%, estimated by running Monte Carlo Markov chains with a length of 20,000 with a burn-in of 10,000 using the Goodman-Weare algorithm.

\section{Results} 
\label{sec:res}

Discriminating between accretion and cooling scenarios for the observed emission from a transient XRP in the quiescent state can be achieved by analyzing both the timing and spectral properties of its emission.

\subsection{Timing analysis}
\label{sec:timing}

\begin{figure}
\centering
\includegraphics[width=0.95\linewidth]{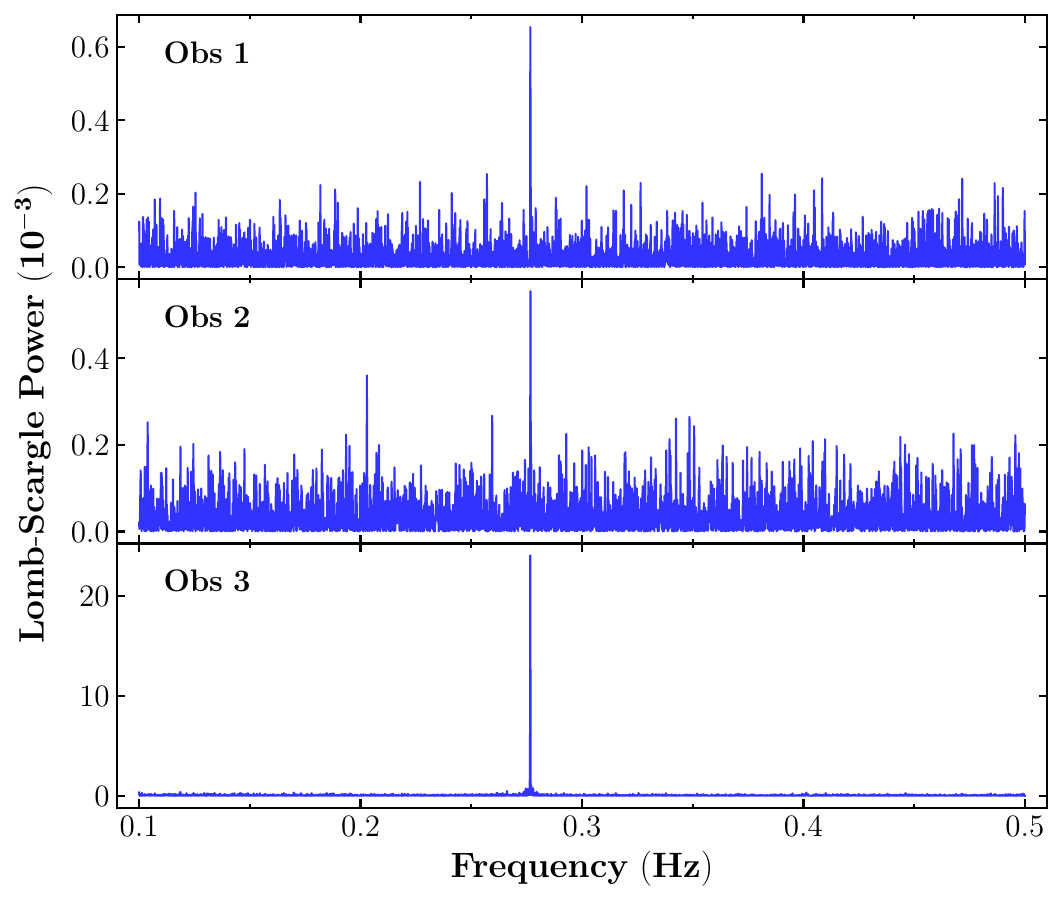}
\caption{Lomb-Scargle periodograms of \mbox{4U 0115+63} derived from three background-subtracted light curves with 0.1\,s bins, observed by {\it XMM-Newton} in 2023. Clear pulsations are detected at a frequency of $\approx$0.277\,Hz, corresponding to the NS spin frequency.}
\label{fig:ls}
\end{figure}

X-ray pulsations from \mbox{4U 0115+63} during its quiescent states have been detected after the 2015 and 2017 giant outbursts \citep{Rouco2017,Rouco2020}. To search for possible periodic signals associated with the NS spin in the data obtained during the 2023 outburst (see Table~\ref{table:obs}), we extracted the background subtracted PN light curves in 0.3--10\,keV band with a bin size of 0.1\,s and calculated their Lomb-Scargle periodograms \citep{Lomb1976, Scargle1982} in 0.1--0.5\,Hz range using the {\sc python} package {\sc astropy}. The results are shown is Fig.~\ref{fig:ls}. We did not use the data from the MOS cameras in the timing analysis due to their poor time resolution (2.6\,s in full frame mode).\footnote{\url{https://heasarc.gsfc.nasa.gov/FTP/xmm/doc/XMM_UHB.pdf}} The Lomb-Scargle periodograms show significant signals at $\approx 0.277$\,Hz in all three observations, corresponding to the NS spin frequency. We examined the confidence levels of these periodic signals, resulting in false alarm probabilities of $2.8 \times 10^{-5}$, $4 \times 10^{-3}$, and $3.2 \times 10^{-209}$ for Obs1, Obs2, and Obs3, respectively. The corresponding pulse periods were precisely estimated using the $Z^2$ method \citep{Buccheri1983} in \textsc{stingray} \citep{Huppenkothen2019} package to be $3.6136 \pm 0.0003\,\rm s$, $3.6143 \pm 0.0003\,\rm s$, and $3.61339 \pm 0.00005\,\rm s$, respectively, where the uncertainties correspond to the standard deviation.

\begin{figure}
\centering
\includegraphics[width=0.95\linewidth]{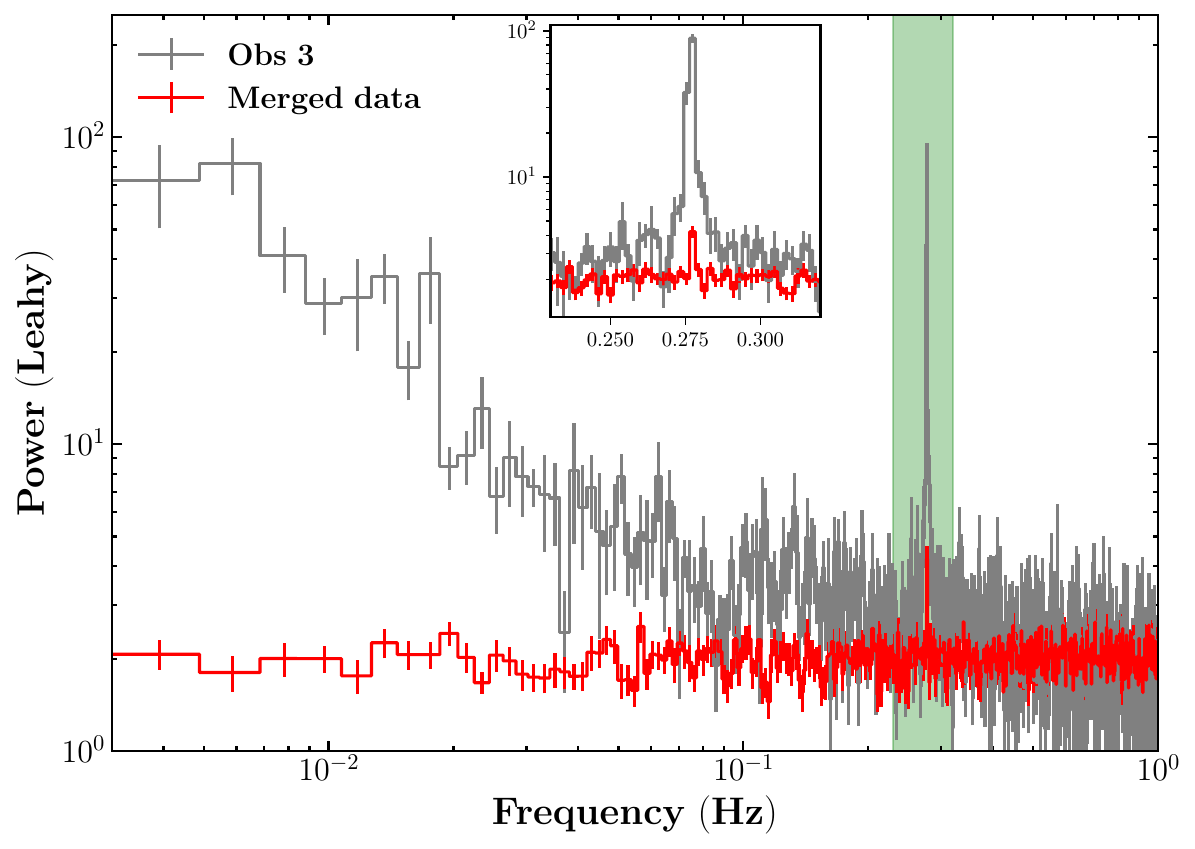}
\caption{PDSs of \mbox{4U 0115+63} normalized as in \citet{Leahy1983} and obtained using 0.1\,s binned, background-subtracted light curves. 
The gray step line represents the PDS from Obs3, which clearly shows a red noise component at low frequencies. The red step line corresponds to the PDS from the combined light curves of Obs1, Obs2, and Obs2016, which exhibits a flat (white noise) distribution in the 0.003--1\,Hz range. The inset provides a zoomed-in view around the NS spin frequency (highlighted by the green stripe).}
\label{fig:PDSs}
\end{figure}

In accreting systems like XRPs, accretion is an intrinsically noisy process, where fluctuations in the mass accretion rate are driven by random variations in the viscosity within the accretion disk \citep{Lyubarskii1997}. Therefore, if accretion is the primary mechanism responsible for the observed X-ray emission, a feasible way to validate this process is by studying the aperiodic variability in the light curves \citep[e.g.,][]{Doroshenko2020}. Specifically, the authors compared the PDSs of accreting pulsars and non-accreting systems (such as magnetars and radio pulsars), showing that low-frequency noise dominates the PDSs of accreting systems. They attributed these differences to the presence or absence of ongoing accretion onto the NS.

To test for the presence of accretion in \mbox{4U 0115+63} during the different states observed with {\it XMM-Newton}, we generated PDSs using the \texttt{powspec} tool from the HEASoft package, as shown in Fig.~\ref{fig:PDSs}.
We note that the {\it XMM-Newton} observation from 2016 (Obs2016) shows an averaged count rate similar to that of Obs1 and Obs2 \citep{Rouco2020}. Therefore, we included this observation in the timing and spectral analysis to improve the statistics.
As clearly seen in Fig.~\ref{fig:PDSs}, the red noise dominates the low-frequency part of the PDS in Obs3. This observation was performed when the NS was near periastron passage (see Fig.~\ref{fig:lc}), a phase where the system is expected to experience an increased accretion rate, potentially leading to a type I outburst. In contrast, the PDS of the merged light curves from Obs1, Obs2, and Obs2016 remains flat at a level of $\sim 2$ at low frequencies, which is characteristic of the white noise. We fitted this PDS with a constant value of 2, yielding a reduced chi-square of $\chi^2_{\rm red} = 0.82$ with 511 degrees of freedom. At the same time, peaks in the PDS corresponding to the spin frequency are significantly detected at $\approx$0.277~Hz in both states. We fitted both peaks with a Gaussian function, resulting in a standard deviation of $3\times 10^{-4}$~Hz for the quiescent observations, whereas the peak in Obs3 appears broadened with a higher standard deviation of $1\times10^{-3}$~Hz, likely due to the interplay between periodic and aperiodic variability.

\begin{figure}
\centering
\includegraphics[width=0.95\linewidth]{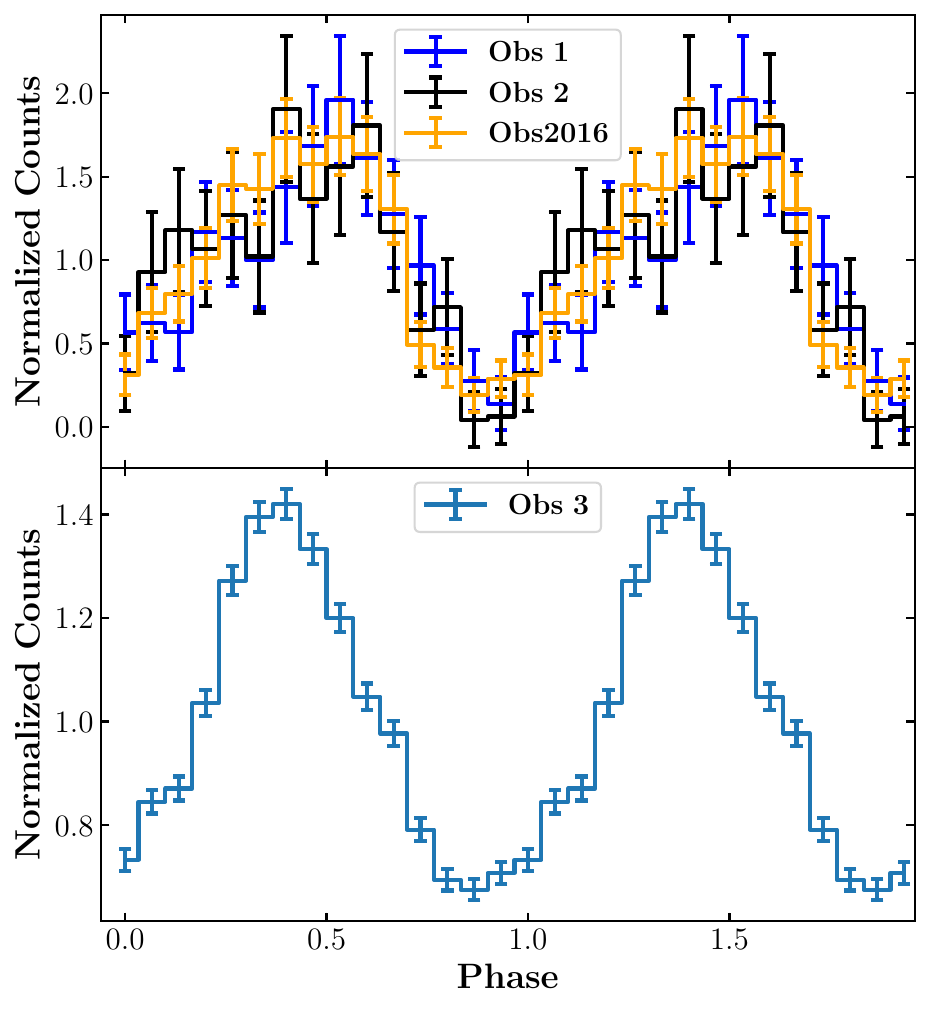}
\caption{Pulse profiles from the \textit{XMM-Newton} observations obtained during the quiescent (top) and accreting (bottom) states of \mbox{4U 0115+63}, folded using background-subtracted light curves with 0.1\,s binning in the 0.3--10\,keV energy range and using spin periods estimated via epoch-folding. 
The profiles were normalized by their average count rates. 
Phases were aligned using the cross-correlation method.}
\label{fig:profile}
\end{figure}

The pulse profiles of \mbox{4U 0115+63} obtained when the source was in the quiescent (Obs1, Obs2, and Obs2016) and accreting states (Obs3) are shown in Fig.~\ref{fig:profile}. For the quiescence, the pulse profiles are very similar, consisting of a sinusoidal-like peak accompanied by a weak harmonic, consistent with the findings of \citet{Rouco2020}. The averaged pulse profile was fitted with a combination of a sinusoid and its first harmonic, revealing an amplitude ratio between the fundamental and harmonic of approximately 3. We quantified the pulse amplitude using the fractional root-mean-square pulsed fraction ($PF$), which is defined as\begin{equation}
    PF = \frac{ \left ({\textstyle \sum_{i=1}^{N}} (r_i -\bar{r})^2/N \right )^{1/2}}{\bar{r}},
\end{equation}
where $N$ represents the number of phase bins, $r_i$ is the count rate in the $i$-th phase bin, and $\bar{r}$ is the phase-averaged count rate. The resulting $PF$ values for Obs1, Obs2, and Obs2016 are $52 \pm 5\%$, $55 \pm 5\%$, and $56 \pm 5\%$, respectively. For comparison, we also calculated the $PF$ for Obs3, which is $26 \pm 6\%$, approximately half the value observed in the quiescent-state observations.

\subsection{Spectral analysis}
\label{sec:spec}

The main aim of the current work is to determine the physical origin (whether the low-level accretion or the NS cooling) of the X-ray emission of the \mbox{4U 0115+63} emission observed in the quiescent state. Therefore, our spectral analysis primarily focuses on the data from Obs1, Obs2, and Obs2016, while Obs3 is used as a reference case with clear evidence of ongoing accretion.

According to observational studies \citep{Tsygankov2016,Wijnands2016,Rouco2017,Rouco2020}, the continuum emission of \mbox{4U 0115+63} in the quiescent states can be described by a blackbody model. We adopted a spectral model consisting of a blackbody component ({\tt bbodyrad in {\sc xspec}}) modified by interstellar photoelectric absorption ({\tt tbabs} in {\sc xspec}). The hydrogen column density, $N_{\rm H}$, was fixed at $9 \times 10^{21}\,\rm cm^{-2}$ following \citet{Wijnands2016}. Element abundances were set to {\tt wilm} \citep{Wilms2000} and photoelectric cross-sections to {\tt vern} \citep{Verner1996}. As shown in Table~\ref{table:para}, the fitting results shows an acceptable goodness-of-fit statistics: C-statistic/dof=86.1/86, 28.1/37, and 229.6/202 for Obs1, Obs2, and Obs2016, respectively. The best-fit blackbody temperatures ($kT_{\rm BB}$)  for these three spectra are $0.44^{+0.02}_{-0.10}$, $0.49\pm0.06$, and $0.43\pm0.02$\,keV, respectively. Assuming isotropic emission and a distance of 5.0\,kpc, the corresponding emission region radii, derived from the model normalization, are $0.39^{+0.07}_{-0.04}$, $0.30^{+0.11}_{-0.05}$, and $0.32\pm 0.03$\,km, respectively. These radii are significantly smaller than the canonical NS radius, suggesting that the X-ray emission originates from localized hot spots at the magnetic poles of the NS.
For Obs3, the spectrum cannot be described by a single {\tt bbodyrad}. We therefore fitted the spectrum using an absorbed power-law model, i.e., {\tt tbabs*(powerlaw)}, which provides a good fit with $\chi^2$/dof= 734.81/732. The resulting parameters indicate a hard photon index of $\Gamma = 0.81\pm0.03$, which is typical of the accreting state.

\begin{figure}
\centering
\includegraphics[width=1\linewidth]{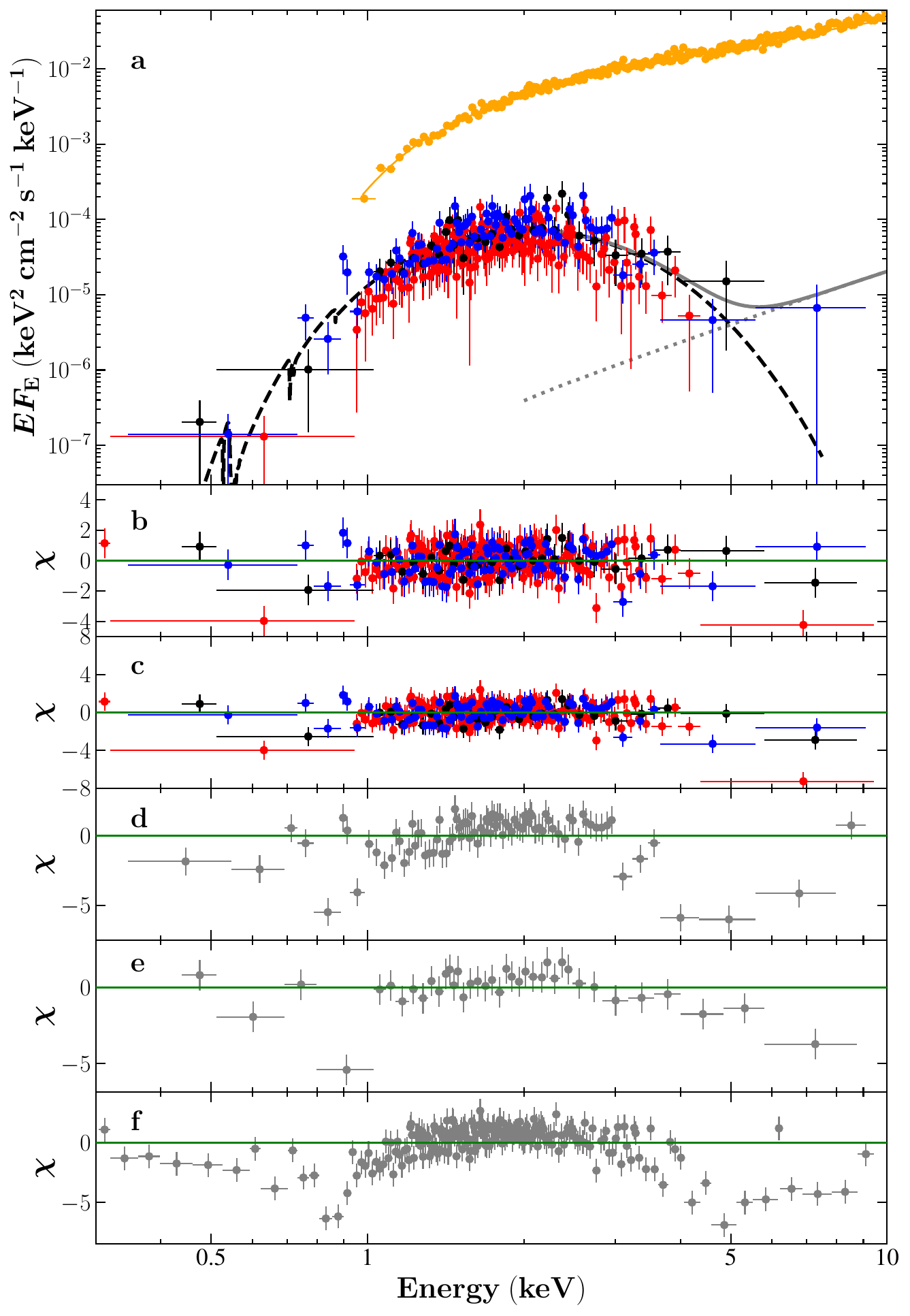}
\caption{Spectral energy distribution of \mbox{4U 0115+63}. 
Panel (a): Unfolded $EF_E$ spectra during {\it XMM-Newton} observations  Obs1 (black), Obs2 (blue), and Obs2016 (red) and fitted jointly. The dashed black curve represents the {\tt bbodyrad} component. The dotted gray %dashed
line shows a possible additional hard component, and the solid gray line indicates the total model spectrum. As a comparison, the spectrum of Obs3 is also in orange, and its low-energy part ($<1\,$keV) is excluded to simplify the panel.  
Residuals from the spectral fits using the {\tt bbodyrad}  and the {\tt bbodyrad + gaussian} models are shown in panels (b) and (c), respectively.
Panels (d), (e), and (f) show the residuals for Obs1, Obs2, and Obs2016, respectively, using the {\tt pegpwrlw} model. 
}
\label{fig:spec}
\end{figure}

We note that the spectral parameters and fluxes of Obs1 and Obs2 are similar to those of Obs2016 \citep{Rouco2017} when modeled with the {\tt bbodyrad} component, indicating that all three observations likely correspond to similar radiative states. Therefore, it is reasonable to fit these spectra simultaneously to improve count statistics. A {\tt constant} component was included to account for possible cross-calibration differences between {\it XMM-Newton} cameras, as well as variations in the source's flux across the different observations. This model provides a good fit to the combined data, with a C-statistic/dof = 343.4/326. The fitting results are shown in Fig.~\ref{fig:spec} and summarized in Table~\ref{table:para}.

At the same time, \citet{Wijnands2016}, \citet{Rouco2017}, and \citet{Rouco2020} suggested that the spectra of \mbox{4U 0115+63} in quiescence can also be described by a power-law model with large photon index. We therefore attempted to replace the {\tt bbodyrad} component with a {\tt pegpwrlw} model in {\sc xspec}. Consistent with \cite{Rouco2020}, the model reveals a soft spectrum with the photon index $\Gamma = 2.7\pm0.2$, $2.4^{+0.3}_{-0.2}$, and  $2.7\pm0.1$ for Obs1, Obs2 and Obs2016, respectively. However, this model did not provide satisfactory fits to the spectra, yielding poor goodness-of-fit values: C-statistic/dof = 173.2/86 for Obs1, 55.4/37 for Obs2 and 464.4/202 for Obs2016. We show the residuals in Fig.~\ref{fig:spec}. 
We also tested other models commonly used for accreting XRPs, such as {\tt comptt} and {\tt cutoffpl}, but the parameters could not be reliably constrained in either case due to the narrow energy band of {\it XMM-Newton}.

Similar to the timing analysis (e.g., the presence of red noise), the spectral shape of an XRP’s emission may provide strong evidence of either ongoing accretion or its absence. Notably, \citet{Tsygankov2019a} demonstrated that even at low mass accretion rates, the X-ray spectrum of XRPs cannot be adequately described by a simple blackbody model. In practice, at low luminosities ($L_{\rm X} < 10^{35}\,\rm erg\,s^{-1}$), XRP spectra typically exhibit a characteristic ``two-hump'' shape. This feature has been reported in several sources, including X~Persei \citep{Doroshenko2012}, GX~304$-$1 \citep{Tsygankov2019a}, 1A~0535+262 \citep{Tsygankov2019b}, GRO~J1008$-$57 \citep{Lutovinov2021}, and 2RXP~J130159.6$-$635806 \citep{Salganik2025}. The hard spectral component is thought to arise partly from cyclotron radiation, originating from interaction of the accretion flow with the NS atmosphere, Comptonized in its upper hot layers \citep{Nelson1995, DiSalvo1998, Coburn2001, Tsygankov2019b, Mushtukov2021, Sokolova-Lapa2021}.

\cite{Mushtukov2021} and \cite{Sokolova-Lapa2021} modeled this scenario and successfully reproduced the overall shape of the low-luminosity spectra of A~0535+262 and \mbox{GX~304$-$1}. Both models predict that the position of the hard component coincides with the cyclotron energy of the pulsar.
To test for the presence of a hard emission component related to the accretion process during the quiescent state, we added a {\tt gaussian} component to our spectral model (i.e., {\tt const*tbabs*bbodyrad}), fixing its central energy at 12~keV, corresponding to the fundamental cyclotron line of \mbox{4U 0115+63} (e.g., \citealt{White1983}). Following the suggestion by \citet{Tsygankov2019a} that the hard component typically contributes about half the flux of the soft component, we constrained the flux of the {\tt gaussian} component in our fitting procedure to be half that of the blackbody component. However, the best-fit results gave a width of the {\tt gaussian} close to zero. This suggests that the hard component is not necessary for the spectral fitting. We further fixed its width at 10~keV to enable a comparison between fits with and without the hard component, resulting a goodness-of-fit of C-statistic/dof=392.2/325. The negative value of $\Delta C=-48.7$ indicates that the two-component model provides a worse fit to the spectrum compared to the single-blackbody model. Therefore, the presence of a hard component in the quiescence spectrum can be tentatively ruled out. The corresponding fits are shown in Fig.~\ref{fig:spec}. However, we note that the limited energy coverage of the available data restricts one to fully characterize the spectra above 10~keV, the spectral results should be interpreted with caution. Nevertheless, due to the low cyclotron line energy of 4U~0115+63, future broadband observations of this source would be valuable for understanding the radiation mechanisms in the low-luminosity state.

\begin{figure} 
\centering
\includegraphics[width=0.95\linewidth]{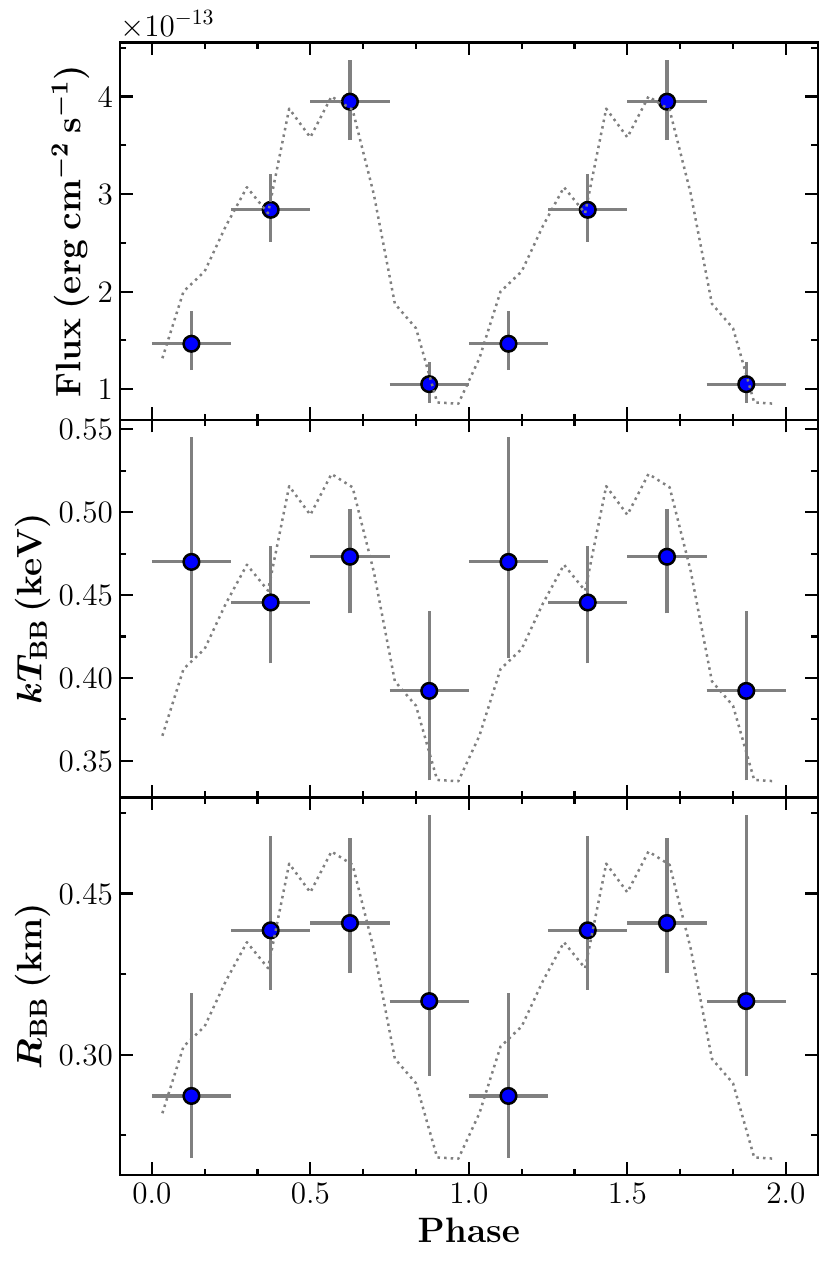}
\caption{Phase-resolved spectral parameters of \mbox{4U 0115+63} obtained using the model {\tt const*tbabs*bbodyrad}. Instrumental cross-calibration constants were fixed at their average values. Unabsorbed X-ray fluxes were calculated in the 0.5–10\,keV energy range. The blackbody radius ($R_{\rm BB}$) is derived from the normalization of the {\tt bbodyrad} component, assuming isotropic emission and a source distance of 5.0\,kpc. The dotted gray lines in each panel indicate the average pulse profiles from Obs1, Obs2, and Obs2016 in arbitrary units. 
}
\label{fig:para}
\end{figure}

Using the single-blackbody model, we performed phase-resolved spectroscopy by dividing the pulse phase and corresponding spectra into four phase intervals ($\phi_{1-4}$ = 0--0.25, 0.25--0.5, 0.5--0.75, and 0.75--1), where phases are aligned with the pulse profiles shown in Fig.~\ref{fig:profile}. The {\tt constant} parameter was fixed to the values obtained from the phase-averaged spectra. The results are presented in Fig.~\ref{fig:para} and Table~\ref{table:para}. The blackbody temperature ($kT_{\rm BB}$) remains relatively stable across most of the spin phase but drops to 0.39\,keV during the pulse trough. 
The blackbody radius exhibits a weak positive correlation with the pulse intensity, varying within the range $R_{\rm BB} =$ 0.26--0.42\,km. However, the low statistics of the phase-resolved spectra are insufficient for further investigating the dependence of flux on $kT_{\rm BB}$ and/or $R_{\rm BB}$.

%%%%%%%%%%%%%%%%%%%%%%%%%
\begin{table*} 
\centering
\renewcommand\arraystretch{1.4}
\caption{Best-fitting spectral parameters for the averaged and phase-resolved spectra of \mbox{4U 0115+63}.}
\footnotesize 
\begin{tabular}{cccccccccc} 
    \hline\hline
      Parameter & Obs1 & Obs2 &  Obs2016 &Averaged & $\phi_1$ & $\phi_2$ & $\phi_3$ & $\phi_4$ \\
    \hline
    $C_{\rm Obs1}$  & - & -  & - &$1.14^{+0.22}_{-0.14}$ & 1.14 (fixed) & 1.14 (fixed) & 1.14 (fixed) & 1.14 (fixed) \\ 
     $C_{\rm Obs2}$ & - & - & - &1 (fixed) & 1 (fixed) & 1 (fixed) & 1 (fixed) & 1 (fixed)  \\
     $C_{\rm Obs2016}$  & - & - & - & $0.73^{+0.13}_{-0.09}$ & 0.73 (fixed) & 0.73 (fixed) & 0.73 (fixed) & 0.73 (fixed)  \\ 
     $N_{\rm H}$ ($10^{22}\,\rm cm^{-2}$)  & 0.9\,(fixed) & 0.9\,(fixed)& 0.9\,(fixed) & 0.9\,(fixed) & 0.9\,(fixed) & 0.9\,(fixed) & 0.9\,(fixed) & 0.9\,(fixed)  \\
     $kT_{\rm BB}$ ($\rm keV$) & $0.44^{+0.02}_{-0.10}$  & $0.49^{+0.06}_{-0.06}$& $0.43^{+0.02}_{-0.02}$ & $0.44^{+0.01}_{-0.01}$ & $0.47^{+0.08}_{-0.06}$ &$0.45^{+0.03}_{-0.04}$ &$0.47^{+0.03}_{-0.03}$ &$0.39^{+0.05}_{-0.05}$   \\  
     $R_{\rm BB}$ (km) & $0.39^{+0.07}_{-0.04}$ & $0.30^{+0.11}_{-0.05}$& $0.32^{+0.03}_{-0.03}$  & $0.36^{+0.04}_{-0.03}$ & $0.26^{+0.10}_{-0.06}$ &$0.41^{+0.09}_{-0.06}$ &$0.42^{+0.08}_{-0.05}$ &$0.35^{+0.17}_{-0.07}$  \\
    Flux ($10^{-13}\rm erg\,cm^{-2}\,s^{-1}$) & $2.38^{+0.15}_{-0.14}$ &  $2.09^{+0.13}_{-0.18}$&1.53$^{+0.01}_{-0.01}$ & $2.10^{+0.15}_{-0.23}$  & $1.47^{+0.34}_{-0.34}$ &$2.84^{+0.37}_{-0.37}$ &$3.95^{+0.43}_{-0.43}$ &$1.05^{+0.22}_{-0.22}$  \\
    $L_{\rm X}$ ($10^{32}\,\rm erg\,s^{-1}$) & $7.3^{+0.5}_{-0.4}$ & $6.4^{+0.4}_{-0.7}$& 4.8$^{+0.3}_{-0.3}$ & $6.4^{+0.5}_{-0.7}$ & $4.5^{+1.0}_{-0.8}$ &$8.7^{+1.1}_{-1.0}$ &$12.1^{+1.3}_{-1.2}$ &$3.2^{+0.7}_{-0.6}$ \\
    \hline
     C-statistic/dof & 86.1/86 & 28.1/37 & 229.6/202 & 343.4/326 & 33.0/29 & 58.5/59 & 86.3/74 & 26.6/26  \\
    \hline
    \end{tabular}
\tablefoot{Unabsorbed X-ray fluxes and luminosities are calculated in the 0.3--10 keV energy range. The blackbody emission radius $R_{\rm BB}$ was estimated by assuming an isotropic source at a distance of 5.0\,kpc.
}
\label{table:para}
\end{table*}
%%%%%%%%%%%%%%%%%%%%%%%%%%
%

%
%%%%%%%%%%%%%%%%%%%%%%%%%%
%%%%%%%%%%%%%%%%%%%%%%%

\section{Discussion and conclusion} 
\label{sec:discussion}

To investigate the origin of the X-ray emission from transient XRPs in the quiescent state, we conducted a timing and spectral analysis of the well-known BeXRP \mbox{4U 0115+63} using {\it XMM-Newton} observations obtained in 2023 following a type II outburst. As in previous post-outburst episodes observed in 2015 \citep{Wijnands2016, Tsygankov2016, Rouco2017} and 2017 \citep{Rouco2020}, the source displayed a low X-ray luminosity of $\sim 10^{33}\,\rm erg\,s^{-1}$ during quiescence, along with type I outbursts occurring near periastron passages.

In systems where the accretion process occurs through the disk, as is expected in \mbox{4U 0115+63}, the red noise arises from the stochastic nature of viscosity within the accretion disk \citep{Lyubarskii1997}. 
Random fluctuations in viscosity lead to local variations in mass density, which propagate both inward and outward through the disk under the influence of viscous diffusion \citep{Mushtukov2018,2019MNRAS.486.4061M}. 
This process generates variability in the mass accretion rate at the inner disk radius, and
the fluctuations are subsequently reflected at the NS surface, resulting in the observed variability of the X-ray flux \citep{Revnivtsev2009}. 
Viscous diffusion acts to suppress mass accretion rate fluctuations at frequencies higher than the local viscous frequency. 
However, even in a cold accretion disk \citep{2017A&A...608A..17T} characterized by low viscosity, initial fluctuations are not entirely damped and can still propagate inward, reaching the inner disk radius and influencing the mass accretion rate onto the NS.
In systems accreting from a stellar wind, variability in the mass accretion rate is instead driven by the intrinsic inhomogeneity, or clumpiness, of the wind \citep{Walter2007}.

Timing analysis revealed the absence of low-frequency red noise in the PDSs of \mbox{4U 0115+63} during quiescence (Obs1, Obs2, and Obs2016). This contrasts sharply with Obs3, obtained near periastron passage, where a prominent red-noise component is detected in the PDS, consistent with the presence of accretion, as also indicated by the spectral analysis. As previously shown by \citet{Doroshenko2020}, the lack of low-frequency variability is a characteristic feature of non-accreting systems, such as magnetars and radio pulsars.

The results of the joint spectral analysis of Obs1, Obs2, and Obs2016 suggest that the continuum emission of \mbox{4U 0115+63} during the quiescent state is well described by a single blackbody model, rather than the two-hump spectrum typically associated with low-level accretion in XRPs. This finding aligns with the timing analysis, which also supports a non-accreting scenario. We therefore conclude that, during these observations, the accreting material did not reach the NS surface due to the action of the centrifugal barrier, i.e., the propeller effect.

At the same time, X-ray pulsations at 0.277\,Hz, corresponding to the NS spin frequency, are clearly detected in this regime in all three observations. The pulse profiles exhibit a sinusoidal-like shape accompanied by a weak harmonic component and a high root-mean-square pulsed fraction ($PF>50\%$), consistent with previous {\it XMM-Newton} observations \citep{Rouco2017, Rouco2020}. This pulsation behavior has been interpreted as evidence of nonuniform thermal emission from the NS surface, likely caused by anisotropic cooling due to the presence of a strong magnetic field in the stellar crust \citep[e.g.,][]{Geppert2006,Geppert2014}. \citet{Perez-Azorin2006} modeled the temperature distribution in the crust of a magnetized NS and found that the resulting X-ray spectrum can be well described by a single blackbody with an emitting area significantly smaller than the full stellar surface, consistent with our results. Depending on the crustal temperature distribution for different model configurations, the predicted pulsed fraction from this cooling scenario is in the range 15--28\% for a NS with  $B\sim 10^{12}$\,G. 

The tentative positive correlation between the blackbody temperature and the X-ray flux over the pulse phase (see the two upper panels in Fig.~\ref{fig:para}) further supports the interpretation of the emission as originating from a cooling atmosphere rather than from ongoing accretion.
In a passively cooling, magnetized NS atmosphere, the temperature typically increases with optical depth \citep[see, e.g.,][]{1992A&A...266..313S}, meaning that observers looking closer to the surface normal see deeper, hotter layers, while those viewing at larger angles sample cooler, upper layers. 
This geometry naturally leads to a positive correlation between the observed flux and blackbody temperature across the pulse phase. 
In contrast, accretion-heated atmospheres are expected to exhibit an inverse temperature profile, with external heating producing hotter upper layers \citep[e.g.,][]{2018A&A...619A.114S}. 
In such a case, the observed flux could be anticorrelated with the apparent temperature, especially in the presence of significant optical depth effects. 
Therefore, the observed tentative temperature–flux correlation may serve as an indicator of anisotropic thermal cooling rather than ongoing accretion. 

On longer timescales, we find that the observed quiescent flux decreased between 2016 and 2023 (see Tables~\ref{table:obs} and~\ref{table:para}). Such variability on year-long timescales cannot be explained by deep crustal heating, which evolves only over $10^4$--$10^5$~yr \citep{Brown1998}. Instead, it points to processes operating in the outer crust. Similar conclusions were reached by \citet{Rouco2017,Rouco2020}, who studied the low-luminosity states of \mbox{4U 0115+63} after its 2015 and 2017 giant outbursts. In that case, the year-to-year differences in flux between post-outburst states can be naturally interpreted as changes in the amount of heat deposited during the preceding accretion episode.

Different authors, based on the magnetic hydrogen atmosphere models of a cooling NS, have theoretically predicted a spectrum similar to a blackbody \citep[e.g.,][]{Zane2000,Ho2001}. We also modeled such emission using undisturbed magnetized hydrogen model atmospheres \citep{Suleimanov2009}. 
We computed a model atmosphere with parameters $k T_{\rm eff} = 0.44$\,keV, 
$M = 1.4\,M_\odot$, $R=$12\,km, and a surface $B$-field normal to the atmosphere of $1.3 \times 10^{12}$\,G.
The emergent model spectrum is described well by a diluted blackbody with the flux $F_E\approx w\,\pi\,B_E(f_{\rm c} T_{\rm eff})$, a color correction factor  $f_{\rm c}\approx 1.25$, and a dilution factor $w\approx 0.41$. This implies that the actual size of the emission region is larger than that inferred from a simple blackbody model, $R \approx w^{-1/2}R_{\rm BB} \approx 1.56\,R_{\rm BB}$.

We also investigated the angular dependence of the emergent model emission and find that radiation emitted tangentially to the surface, i.e., along the atmosphere plane, is less shifted to high energies  ($f_{\rm c} \approx 1.1$ for the zenith angle, $\theta$, of 81\degr). It is interesting that the ratio of the observed blackbody temperatures of the pulse-average flux spectrum and the low-flux spectrum of the pulse profile is approximately equal to the ratio of the color correction factors for the flux spectrum and the intensity spectrum emitted at $\theta=81\degr$: $0.44\,{\rm keV}/0.39\,{\rm keV}\approx 1.25/1.11 \approx1.13$. 

This means that the observed decrease in blackbody temperature can be explained by the model in which, at the pulse phase with the lower temperature, we see the bright spot inclined at a large angle to the line of sight. Thus, we conclude that a model of the magnetized NS atmosphere can explain the observed spectrum of the source and its phase dependence. More detailed modeling will be presented in a separate paper.

Future observational and theoretical studies of XRPs across different luminosity states will be valuable for advancing our understanding of emission mechanisms in highly magnetized NSs in quiescence. In particular, measurements of polarization properties could reveal the pulsar orientation \citep[see, e.g.,][]{Poutanen2024b} and help verify the models proposed above.

\begin{acknowledgements}
HX acknowledges support from the China Scholarship Council (CSC). 
AAM thanks UKRI Stephen Hawking fellowship.
This research has made use of software provided by the High Energy Astrophysics Science Archive Research Center (HEASARC), which is a service of the Astrophysics Science Division at NASA/GSFC and the High Energy Astrophysics Division of the Smithsonian Astrophysical Observatory. This work is based on observations obtained with {\it XMM-Newton}, an ESA science mission with instruments and contributions directly funded by ESA Member States and NASA.
\end{acknowledgements}

\bibliographystyle{aa}
\bibliography{allbib}

\end{document}